\documentstyle[psfig]{l-aa}
\begin{document}
\thesaurus{08.02.1; 08.14.1: 0816.7 4U1907+09; 13.25.5}
\title{Detection of a cyclotron line and its second harmonic in 4U1907+09} 
\author{G. Cusumano\inst{1}, T. Di Salvo\inst{2}, L. Burderi\inst{2}, 
M. Orlandini\inst{3}, S. Piraino\inst{1}, N. Robba\inst{2}, A. Santangelo\inst{1}}
\institute{Istituto di Fisica Cosmica ed Applicazioni all'Informatica C.N.R.,
Via U. La Malfa 153, I-90146, Palermo, Italy \and Istituto di Fisica 
dell'Universit\`a -- Via Archirafi 36, 90123 Palermo, Italy \and 
Istituto Tecnologie a Studio Radiazioni Extraterrestri, C.N.R., Via Gobetti,
101, 40129 Bologna, Italy}
\offprints{G. Cusumano: cusumano@ifcai.pa.cnr.it}
\date{Received ....; accepted ....}
\maketitle
\markboth{G. Cusumano {\it et al.}: Detection of a cyclotron line and its 
second harmonic in 4U1907+09  }
{G. Cusumano {\it et al.}: Detection of a cyclotron line and its second 
harmonic in 4U1907+09 }

\begin{abstract}
We report the detection of a cyclotron absorption line and its second harmonic 
in the average spectrum of the high mass X-ray binary 4U1907+09 observed by 
the BeppoSAX satellite on 1997 September 27 and 28.
The broad band spectral capability of BeppoSAX 
allowed a good determination of the continuum against which the two 
absorption features 
are evident at $\sim$ 19 and $\sim$ 39 keV.
Correcting for the gravitational redshift of a $\sim 1.4$ M$_{\odot}$ neutron
star, the inferred surface magnetic field strength is $\mathrm B_{surf} =
2.1 \times 10^{12}$ G that is typical for this kind of systems. 
We also detected an iron emission line at $\sim 6.4$ keV, with 
an equivalent width of $\sim$ 60 eV.

\keywords{ binaries: close --- star: neutron --- pulsars: individual: 4U 1907+09 
--- X--rays:stars}
\end{abstract}

\section{Introduction}
 4U1907+09 is an X-ray pulsar powered by wind accretion from a close blue 
 supergiant companion star.
  It belongs to the class of the high-mass X-ray binaries (see, e.g., White, 
  Nagase $\&$ Parmar 1995).\\
 4U1907+09 was discovered in the
Uhuru surveys (Giacconi {\it et al.} 1971, Schwartz {\it et al.} 1972).
An observation of this source performed by the RXTE satellite has
been recently analyzed by in 't Zand {\it et al.} (1998).
The reported pulse period is $ 440.341 \pm 0.006$ s.
No pulsations were found above 40 keV up to date.
The reported orbital period is $8.3753 \pm 0.0001$ d
and the eccentricity is $0.28 \pm 0.04$.
The pulsar showed a fairly constant spin-down between 1983 and 1996 with
a mean derivative of $\mathrm \dot{P}= +0.225$ s yr$^{-1}$.
Periodic flares, two per orbit, have been reported
(Marshall $\&$ Ricketts 1980) with an increase of the intensity by about a
factor 5.
The energy spectrum of 4U1907+09 was observed with several X-ray missions 
(Ariel V, Tenma, EXOSAT, Ginga, XMPC and RXTE) covering, not simultaneously, 
the range 2--100 keV. 
It was fitted by an absorbed power law (N$_H$ in the range
$1.5 - 5.7 \times 10^{22}$ cm$^{-2}$ and photon index in the range
0.83 - 1.52 -- see Schartz {\it et al.} 1980, Marshall and Ricketts 1980,
Makishima {\it et al.} 1984, Cook and Page 1987, Chitnis {\it et al.} 1993).
Data from a Ginga observation of 4U1907+09 were fitted using
the NPEX model (Negative and Positive power-laws EXponential)
as continuum plus
a Lorentzian shaped cyclotron absorption line at $\sim$ 19 keV and an iron 
line at  6.6 keV with a flux of 1.7$\pm$0.5 ph s$^{-1}$.
In the RXTE observation the spectrum of 'non dipping' data, subtracted by the
'dipping' spectrum, was fitted in the 2-30 keV range
by an absorbed power law with high energy cut--off at 13.6 keV
with no evidence of any absorption features (in 't Zand {\it et al.} 1997).
\\
In this paper we report the results of a spectral analysis of the broad band
spectrum (1--80 keV) of 4U1907+09 observed with the BeppoSAX Narrow Field
Instruments (NFIs). We confirm the presence of the absorption feature at
$\sim 19$ keV and we report the detection of a strong second harmonic.
 
\section{Observation}
4U1907+09 was observed by the 
NFIs on board BeppoSAX (\cite{boe_1})
on 1997 September 27 and 28.
The NFIs consist of four coaligned instruments: the Low Energy
Concentrator Spectrometer (LECS) operating in the energy range
0.1--10~keV (\cite{par}),
the Medium Energy Concentrator Spectrometer (MECS) having three
units operating in the range 1--10~keV (\cite{boe_2}), the High Pressure
Gas Scintillation Proportional Counter (HPGSPC) operating in the range
4--120~keV (\cite{man}) and the Phoswich Detector System (PDS) with four
scintillation units operating in the range 15--300~keV (\cite{fro}).
The energy resolution ($\Delta$E/E) is 8\% at 6 keV, 6\% at 20 keV and 
12$\%$ at 40.0 keV.
We do not consider LECS data because the MECS statistics is more than enough
to model the continuum under the cyclotron lines.
The total exposure time was 68576 s for the MECS,
33079 s for the HPGSPC and 34731 s for the PDS.
HPGSPC and PDS were exposed about half as long as the MECS
due to the rocking collimator observation mode.
The MECS events were reduced using the SAXDAS v.1.1.0 package (BeppoSAX 
Cookbook, http://www.sdc.asi.it/software/cookbook) while HPGSPC and PDS data 
were reduced using the XAS v.2.0.1. package (Chiappetti \& Dal Fiume 1997).
Response matrices were used which were made publicly available on August 31,
  1997.

The 4U1907+09 background subtracted light--curves of the three NFIs 
are reported in Figure 1. The gaps in the light curves are due to
non--observing intervals during 
passes over 
  the South Atlantic Anomaly and during Earth occultations.
An increase in intensity of about a factor two (up to 13
c/s, $\sim 45$ mCrab in the MECS) is observed at $\sim 2500$ s
after the beginning of the observation. 
Using the $P_{orb}$ reported by in 't Zand
{\it et al.} (1998), the flux enhancement observed by BeppoSAX is found to occur
$613.93 \pm 0.05$ orbital periods after the primary flare observed by Tenma
(Makishima {\it et al.} 1984). Therefore we tentatively identified this episode
with the periodic primary flare, 
confirming the occurrence of the phase-locked flares.

\begin{figure}
\centerline{
\hbox{
\psfig{figure=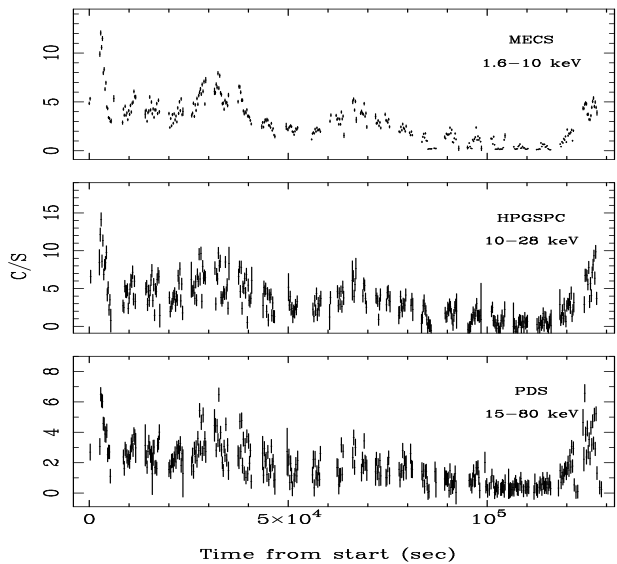,height=7.7cm,rheight=7.5cm,width=9cm,clip=}}}
\caption{Light--curves of the NFIs binned at the spin period of 4U1907+09}
\label{fig1}
\end{figure}

\section{Spectral Analysis}
Spectral analysis was performed on the spectrum averaged over the complete observation.
Due to the presence of the galactic ridge and the supernova remnant W49B 
($\sim 0.8^{\circ}$ from 4U1907+09) additional care has been applied in the 
determination of the background.
The local background has been measured in a region of the image 
20 arcminutes away from the source for the MECS, and 
compared with the background of the archive blank fields. The MECS
local background spectrum can be well represented by the blank field increased
by a factor 1.6. Therefore blank sky background has 
been used for the MECS, applying the above correction.
For the collimated instruments we use the local background accumulated in the 
off--source position. Anyway, from previous measurements of 
W49B (\cite{Fuj}, Smith et al. 1985) and the galactic ridge (Yamasaki et al. 
1997) and 
considering the  strong reduction in the effective area for off axis sources
we estimate that the contribution of these sources 
is negligible above 10 keV.

The energy range used in the spectral analysis for 
each NFI was: 1.6--10 keV for MECS, 10--30 keV for 
HPGSPC and 15--80 keV for PDS. 
The continuum in the range 1--80 keV has been modeled adopting
an absorbed power law with an exponential cut--off for energies above 
E$_{\mathrm cut}$. 
To take into account the systematic differences in the normalization 
between the different NFIs (Cusumano {\it et al.} 1998) the relative 
normalizations have been kept as free parameters. 
The  $\chi_{red}^2$ value
with the above continuum model was 2.14 (221 dof).
Figure 2 shows the spectra with 
the best fit continuum and the residuals.

\begin{figure}
\vbox{
\psfig{figure=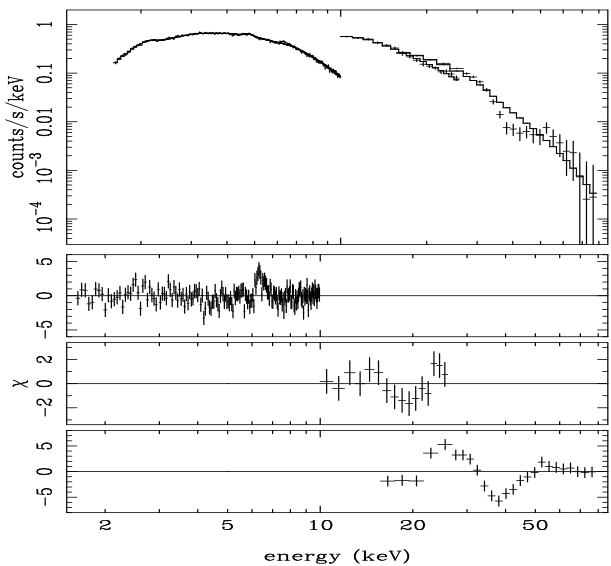,height=7.8cm,width=9cm,clip=}}
\caption{1.6--80 keV spectrum of 4U1907+09 with the best fit continuum 
model (top) and residuals (separately for each NFI in the bottom three 
panels).}
\label{fig2}
\end{figure}

\begin{figure}
\vbox{
\psfig{figure=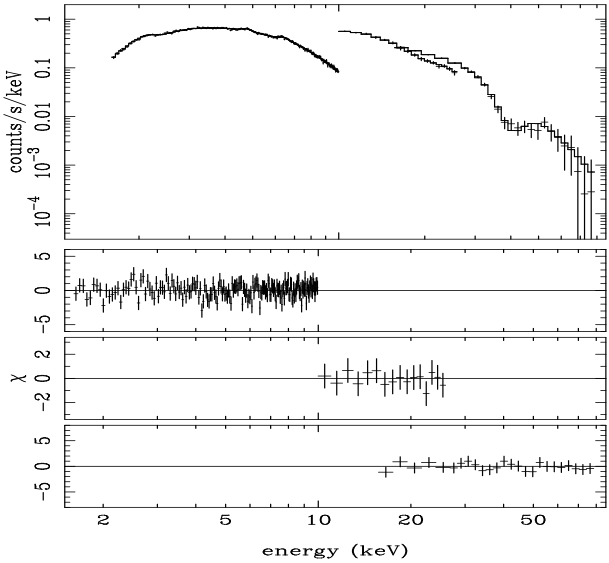,height=7.8cm,width=9cm,clip=}}
\caption{1.6--80 keV spectrum of 4U1907+09  with the best fit continuum model
plus an emission gaussian line at $\sim 6.4$ keV and two absorption gaussian
lines at 18.8 and 39.4 keV. The residuals do not show significant features.}
\label{fig3}
\end{figure}

The excess at $\sim 6.4$ keV in the residuals strongly
suggests the presence of an iron emission line.
Therefore we included  an additional gaussian  to model the iron line.
\\
Other features are evident from the residuals above 10 keV.
A feature at $\sim$ 40 keV is evident.
We tried several models to fit this feature and we obtained the best fit
including an absorption according to a gaussian shaped 
feature (Soong et al. 1990) centered at  $\sim 39$ keV.
The $\chi_{red}^2$ value after the addition of the absorption
gaussian feature was 1.27 (215 dof).
A feature at $\sim$ 19 keV is also present in the residuals of Fig.
2. We tried to improve the fit adding an absorption gaussian factor 
to the above model (model 1).
The $\chi_{red}^2$ value after addition of this last gaussian shaped feature was
1.06 (212 dof).
Figure 3 shows the result of the fit. The residuals do not show significant
structures.
The F-test for three additional parameters gives a significance 
of the absorption feature at $\sim$19 keV greater than 99.999\%. 
The significance of the last feature and its energy that is compatible (Table
 1) with half of the energy of the other absorption feature allow us to identify
these two
absorption features as the
fundamental and the second harmonic of the cyclotron absorption line
at $\sim 19$ keV reported by
Makishima and Mihara (1992) from {\it Ginga} observations.
The Lorentzian absorption model of Mihara
(1995) (model 2) was also tried to fit the two absorption features.
Also in this case the fit gives an acceptable $\mathrm \chi^2$.
Best fit parameters obtained with the models described above are reported in 
table 1.
\\
Other descriptions of the continuum (such as NPEX, Mihara 1995), together
with the models for cyclotron lines described above, gave worse results. 
\\
The luminosity of the source during the BeppoSAX observation was 
$4\times10^{34}D^2_{kpc}$ in the 2--10 keV energy range and 
$10^{35}D^2_{kpc}$ in the 2--80 keV range where $D^2_{kpc}$, the distance 
of 4U1907+09 in kpc, has been roughly estimated between 2.4 and 5.9 kpc 
(Van Kerkwijk et al. 1989). 

\begin{table}
\caption[]{best fit parameters}
\begin{flushleft}
\begin{tabular}{llll}
\hline\noalign{\smallskip}
\noalign{\smallskip}
             & model 1          & model 2         &    \\
N$_H$     & 2.81 $\pm$ 0.04   & 2.81 $\pm$0.04    &    \\
$\alpha$  & 1.27 $\pm$ 0.01   & 1.27 $\pm$0.01    &    \\
Norm      & 0.052 $\pm$ 0.001 & 0.052$\pm$0.001   &    \\
E$_{iron}$ (keV) & 6.47 $\pm$ 0.03 & 6.47 $\pm$0.03  &   \\
$\sigma_{iron}$ (keV) &  $\leq 0.15$ &  $\leq 0.15$      &  \\
I$_{iron}$  & $(3.0 \pm 0.7) \times 10^{-4}$ & (3.0$\pm$0.7) $\times$ 
10$^{-4}$ & \\
E$_{cut}$ (keV)  & 12.0 $\pm$ 0.3  & 12.3 $\pm$0.4    &  \\
E$_{fold}$ (keV) & 12.0 $\pm$ 0.3 & 13.8$\pm$0.8     &  \\
E$_{CRF1}$ (keV) & 18.8 $\pm$ 0.4       & 19.3$\pm$0.2 &  \\
$\sigma_{CRF1}$ (keV) & 2.2 $\pm$ 0.4       &     &    \\
EW$_{CRF1}$ (keV) & 2.3 $\pm$ 0.4      &     &    \\
$D_{CRF1}$        &                    & 0.29$\pm$0.05 & \\
W$_{CRF1}$ (keV)  &                    & 1.8$\pm$0.8  & \\
E$_{CRF2}$ (keV)  & 39.4 $\pm$ 0.6 & 2 $\times$ E$_{CRF1}$ & \\
$\sigma_{CRF2}$ (keV)  &  3.6 $\pm$ 0.7    &              &     \\
EW$_{CRF2}$ (keV) &  16.7 $\pm$ 2.1  &         &     \\
$D_{CRF2}$        &                        & 2.7$\pm$1.1   &  \\
W$_{CRF2}$ (keV)  &                        & 2.8$\pm$1.3 &  \\
$\chi^2_{rid}$ (d.o.f.) & 1.06 (212)  & 1.03(213) & \\
\noalign{\smallskip}
\hline
\multicolumn {4} {l} {Norm is in unit of  photons/keV/cm$^2$/s a 1 keV.}    \\
\multicolumn {4} {l} {N$_H$ is in unit of ($10^{22}$ atoms cm$^{-2}$).}   \\
\multicolumn {4} {l} {I is the total photons cm$^{-2}$ s$^{-1}$ in the line.}\\
\multicolumn {4} {l} {Quoted errors refer to single-parameter 68\% confidence level}\\ 
\multicolumn {4} {l} {The upper limit to the $\sigma_{iron}$ is at 90\% confidence level.} \\
\end{tabular}
\end{flushleft}
\end{table}

\section{Discussion}
The detection of cyclotron lines in the hard X-ray spectrum of binary 
pulsars 
allows a direct measurement 
of the magnetic field of strong magnetized neutron stars.
{\it Ginga}, with its relatively broad energy band (1-40 keV), allowed 
the detection of cyclotron absorption lines in several X-ray pulsars.
Sometimes also the presence of a second harmonic was claimed. 
However the detection of these harmonics was often ambiguous, 
as in the case of 4U1538-52 (Clark et al. 1989). In fact only the
red wing of the second harmonic (at $\sim 40$ keV) could be seen 
with {\it Ginga} and a high energy cut-off could also fit the spectrum.
The very broad energy band of BeppoSAX NFIs allows to measure the X-ray
spectrum of 4U1907+09 from 1 to 80 keV, well above the top end energy of
Ginga Large Area Counter. So we can constrain the shape of the
underlying continuum with great accuracy, allowing an easier measure
of spectral features, when present. In the case of this source, we detect
two absorption features at 18.8 keV and 39.4 keV respectively, 
which we interpret as two cyclotron lines, the fundamental and the 
second harmonic.
In this case one might expect that the line at 19 keV would be deeper than
the line at 39 keV because the cyclotron opacity decreases for higher 
harmonics. On the contrary for this source the second line appears deeper
than the first line. A similar line structure was found in the spectrum of 
GB 880205 (e. g. Alexander and Meszaros 1989). According to Alexander and 
Meszaros (1989) this line structure can be explained through the process
of Compton scattering with multiple photon emission. In fact, in the multiple
emission case, each scattering removes a photon with $E \sim 2 E_c$ and
gives two photons with $E \sim E_c$. Thus photons at $\sim 2 E_c$ are depleted,
while those at $\sim E_c$ replenished, and the line at the second harmonic
becomes deeper than the fundamental.

Considering the relation between the cyclotron energy and the magnetic field 
$E_c$/(1 keV) = 11.6 B/($10^{12}$ Gauss), the observed value of the cyclotron 
energy of 18.8 keV implies a magnetic field of $B_{obs} \simeq 1.6 \times
10^{12}$ Gauss.
If the cyclotron absorption takes place near
the neutron star surface, where the magnetic field is strong, the observed 
resonance energy will be affected by the gravitational redshift:
$E_c^{obs} = E_c (1+z)^{-1}$, with 
\begin{equation}
(1+z)^{-1} = \left(1 - \frac{2 G M_{NS}}{R c^2} \right)^{1/2}
\end{equation}
where $M_{NS}$ is the mass of the neutron star and R is the distance of
the region where the line is formed from the center of the neutron star. Using
$M_{NS} = 1.4 M_{\odot}$ and $R = 10^6$ cm, we get $(1+z)^{-1}$ = 0.76 and
$E_c \simeq 24.7$ keV. In this case the magnetic field should be
$B_{{\rm surf}} \simeq 2.1 \times 10^{12}$ Gauss.

An estimate of the temperature of the region in which the
cyclotron absorption is formed can be obtained from the width of the observed
lines. In fact if the electrons have a temperature $kT$, the energy of the
scattered photons distributes in a gaussian-like peak with 
Lorentzian wings (Voigt function).
In this case the doppler broadening of a line $\Delta E / E$ is constant and 
is given by (see {\it e.g.} Rybicki and Lightman 1979):
\begin{equation}
\Delta E = E_c \sqrt{\frac{2kT}{m c^2}}
\end{equation}
a factor $\cos\theta$ -- where $\theta$ is the angle between the magnetic
field lines and the line of sight -- arises if the plasma motion is confined
by the field. Averaging this over angles between 0 and $\pi/2$ introduces 
corrections $\sim 1$.
We observe that the obtained width of the second harmonic is double that of
the fundamental, in agreement with the relation above. 
The obtained values of the parameters give a temperature of 4 -- 7 keV.\\

\end{document}